\documentstyle[prb,aps,preprint,tighten]{revtex}

\begin{document}
\draft
\preprint{CTH/GU-ITP-94-2}

\title{\bf Disorder Effects in Two-Dimensional d-wave Superconductors}
\author {A. A. Nersesyan\cite{email}}
\address{
Institute of Theoretical Physics,
Chalmers University of Technology, S-41296 G\"{o}teborg, Sweden
\\
and Institute of Physics,
Georgian Academy of Sciences, Tamarashvili 6, 380077, Tbilisi, Georgia}

\author{A. M. Tsvelik}
\address{
Department of Physics, University of Oxford, 1 Keble Road, Oxford, OX1 3NP, UK}
\author{F. Wenger}
\address{
Institute of Theoretical Physics,
Chalmers University of Technology, S-41296 G\"{o}teborg, Sweden}
\date{\today}
\maketitle
\begin{abstract}
Influence of weak nonmagnetic impurities on the single-particle
density of states $\rho(\omega)$
of two-dimensional
electron systems with a conical spectrum is studied.
We use a nonperturbative approach, based on replica
trick with subsequent mapping
of the effective action onto a one-dimensional
model of interacting fermions,
the latter being treated by Abelian and non-Abelian
bosonization methods.
It is shown that, in a d-wave superconductor, the density of states, averaged
over randomness, follows a nontrivial power-law behavior near the Fermi energy:
$\rho(\omega) \sim |\omega|^{\alpha}$. The exponent $\alpha>0$ is calculated
for several types of disorder. We demonstrate that the property
$\rho(0) = 0$ is a direct consequence of a {\it continuous} symmetry
of the effective fermionic model, whose breakdown is forbidden
in two dimensions. As a counter example, we consider another model with a
conical
spectrum - a two-dimensional orbital antiferromagnet, where static disorder
leads to a finite $\rho(0)$ due to breakdown of a {\it discrete}
(particle-hole) symmetry.
\end{abstract}
\narrowtext
\section{Introduction}

The problem of weak static disorder in electronic systems with extended Fermi
surface has been extensively studied \cite{Lee85}. It is well understood that
randomness has a profound influence on the transport properties and weakly
affects the density of states. The latter is no longer true, if the
density of states (DOS) of a pure system vanishes linearly
at zero energy: $\rho(\omega) \sim |\omega|$. In this case the standard
procedure of averaging over impurities is complicated by the
appearance of logarithmic singularities in the perturbative expansion of the
single-electron self energy \cite{Gorkov85,Ueda85,Gross86,Kusma85}.
$$
\Sigma(\omega_n) = i\omega_n( - g\ln(\Lambda/|\omega_n|) + ... ).
$$
Such a situation
 can be realized in two-dimensional systems with a Dirac-like, conical
spectrum $E^2(p) = v_x^2 p_x^2 +  v_y^2 p_y^2$, describing low-energy states
near a degeneracy point (node), as well as in three-dimensional
polar superconductors, where the order parameter has a nodal line. In
both cases the pure systems exhibit a $T^2$ low temperature
specific heat. Since the exponents of thermodynamic quantities are used  as
an experimental  criterium for selection of possible order parameters, it is
important to study their stability with respect to disorder. For 3D
polar superconductors, the problem is simplified by the fact that,
due to the finite size of the Fermi surface,
the diagrams with crossing impurity lines are not divergent,
and one can easily sum the remaining diagrammatic series. This was done by
Gor'kov and Kalugin \cite{Gorkov85}, who showed
that weak nonmagnetic disorder gives rise to
a special energy scale $Q_0 \sim \Lambda \exp[ - const/c]$,
$c$ being the impurity concentration, below which the
DOS becomes finite $\rho(0) \sim Q_0$.

The renewed interest in quenched disorder in 2D systems$^{2,3}$
has been stimulated by a number of
experimental indications that the pairing state in the
copper-oxide superconductors
may be of $d_{x^{2} - y^{2}}$ symmetry.
In a recent publication, Lee \cite{Lee93} applied the
results of Ref.\onlinecite{Gorkov85} to analyze the role of nonmagnetic
impurities
in a 2D d-wave superconductor. He argued that, since a system with
a nonzero DOS formally resembles a metal with a finite Fermi surface,
the standard scaling arguments \cite{Lee85} would indicate localization of
all low-energy states.

In this paper representing an extended version of our short report
\cite{Ners93},
 we show that,  when averaging over impurities
in 2D systems with a {\it finite} number of isolated nodes,
the logarithmic singularities appear in {\it all } self-energy diagrams,
including the crossing ones. No specific subset
of diagrams can be selected from the whole self-energy series expansion.
What one finds here is a typical "logarithmic" situation
when logarithmic singularities appear in  all
self-energy diagrams, including the crossing ones.
Approximations based on partial summation of the ("rainbow") diagrams
\cite{Gorkov85,Gross86} do not apply.
We overcome this difficulty by means of the field-theoretical approach based
on replica trick. The static nature of the disorder allows to represent
the effective, disorder-free action in terms of a one-dimensional
model of interacting fermions. Using then bosonization technique,
we demonstrate that, instead of creating a finite DOS at $\omega = 0$,
in the 2D d-wave superconductor the disorder changes the exponent
of the DOS: $\rho(\omega) \sim |\omega|^{\alpha},$ $(0 < \alpha < 1)$.
The magnitude of $\alpha$ depends on the type of disorder.
For a slowly varying (quasiclassical) random potential, when only forward
scattering is present, $1 - \alpha \sim c$. The internode backscattering
processes strongly reduce  $\alpha$ at energies
$\sim Q_0$, making it concentration independent: $\alpha  = 1/7$.
The picture indeed becomes very close to the one with a finite DOS, but
still not the same. In this situation, the renormalization group for the
conductivity can lead to an intermediate fixed point, which suggests a
finite conductivity.

Superconductors are not the only systems where a conical spectrum and,
as a consequence, linear DOS can appear.
Such a spectrum occurs in zero-gap (degenerate) semiconductors
\cite{Fradkin86},
and in heterojunctions
where the contact is made between  semiconductors with inverted
symmetry of bands \cite{Volkov85,Kusma85}.
It can be realized in 2D graphite sheets \cite{Semen84}, as well
as for lattice electrons in a strong magnetic field, the most well
known example being a tight-binding model of fermions with
1/2 of a magnetic flux quantum per plaquette (flux phase) \cite{Fisher85}.
The conical spectrum is also a property of hypothetical
orbital antiferromagnet and spin nematic states \cite{Ners88}.

Despite the common feature of all these systems, - the existence of the
degeneracy points (nodes) in their spectrum, the role
of disorder can be quite different in each
particular case, depending on the symmetry of the pure system.
 The result is determined by the structure of random terms
which appear in the effective massless Dirac Hamiltonian \cite{Fradkin86},
describing low-energy states of the pure system.
In the 2D d-wave superconductor, nonmagnetic impurities
add to the Dirac model random gauge fields \cite{Lee93}, abelian or
non-abelian, depending on the type of included scattering processes.
The random gauge fields do not break particle-hole symmetry in the
vicinity of the nodes and lead to a critical, power-law
behavior of $\rho(\omega)$ at $\omega \rightarrow 0$.
As we demonstrate below, this is a direct consequence
of the fact that at $\omega \rightarrow 0$
the equivalent 1D fermionic model possesses
{\it continuous} replica and chiral $\gamma^5$ symmetries, remaining unbroken
in two dimensions.

On the other hand, in 2D systems with a particle-hole condensate,
such as the flux phase or orbital antiferromagnet, the situation is
different. The nonmagnetic disorder gives rise to a random
chemical potential and random charge-density wave, the latter being
equivalent to a random relativistic mass of the Dirac quasiparticles.
We argue that the random Dirac mass alone would also lead to a
critical behavior with $\rho(0) = 0$. However, the random
chemical potential, which is always present even for a weak quasiclassical
disorder, breaks the particle-hole symmetry and drives the system away
from criticality. The special role of the random chemical potential is that it
lowers the continuous chiral symmetry of the corresponding 1D Fermi model
down to a {\it discrete} one. The spontaneous breakdown of the latter
is signalled by a finite $\rho(0)$, as previously pointed
out by Fisher and Fradkin \cite{Fisher85}. It is worth noticing that,
for a d-wave superconductor, the appearance of a random chemical potential
in the effective Dirac model could be only caused by random magnetic fields.

The paper is organized as follows. In Section II we formulate a model
of two-dimensional d-wave superconductor with non-magnetic
impurities. In Section III we give a perturbative analysis of
the disorder in terms of the diagram expansion for
the single electron Green's function. The expansion demonstrates
that the number of singular diagrams in the given $n-$th order
grows with $n$, and the problem is non-perturbative. In Section IV
we average over the randomness using the replica trick and
reformulate the replicated model as a two-dimensional Euclidean
relativistic field
theory. In Section V we consider a simplified version of this theory,
namely, we consider only scattering within each node. This simplified
theory is solved by the ordinary bosonization. In Section VI we
consider the scattering between nodes; the corresponding model is
solved by the non-Abelian bosonization. In Section VII we describe
another model with a conical spectrum - a 2D orbital
antiferromagnet. We claim that  this model has a finite $\rho(0)$.
The paper has an Appendix where we discuss some properties of the
Bethe-ansatz solution of the replicated model of impure d-wave
superconductors.

\section{A Model of Impure d-wave Superconductor}

 We start from the following model describing low energy electronic
degrees of freedom in a 2D d-wave superconductor:
\begin{equation}
H =
\sum_{{\bf k},\sigma}
\epsilon({\bf k}) c^+_{{\bf k}\sigma}c_{{\bf k}\sigma} -
\sum_{\bf k}[\Delta({\bf k})c^+_{{\bf k}\uparrow}c^+_{{\bf k}\downarrow}
+ H. c.]. \label{eq:Ham}
\end{equation}
Here
$\epsilon({\bf k})$ is a tight-binding spectrum in the normal state
which respects full point
symmetry of the underlying (square for the copper oxides) lattice, and
$\Delta({\bf k})$ is a pairing amplitude, odd under
$\pi/2$-rotations in the ${\bf k}$-space (the $d_{x^2 - y^2}$ symmetry).
Without loss
of generality we can choose
$\epsilon({\bf k}) = -2t(\cos k_x + \cos k_y)-\mu$ and
 $\Delta({\bf k}) = \Delta_0(\cos k_x - \cos k_y)$. It will be assumed that
$\Delta_0 \ll t $.
The  quasiparticle
spectrum $E{({\bf k})} = \pm \sqrt
{\epsilon^{2}({\bf k}) + \Delta^{2}({\bf k})}$ has four nodes at the
Fermi level which we denote $1, \bar 1, 2, \bar 2$ (see Fig. \ref{figure1}).
Their positions are
${\bf k}_1 = - {\bf k}_{\bar 1} = (k_0, k_0);~ {\bf k}_2 = - {\bf k}_{\bar 2}
= ( - k_0, k_0)$, where $k_0 = \arccos(\mu/4t)$. Associated with these nodes
are  gapless excitations with formally relativistic (conical) spectrum at
$|\omega| \ll \Delta_0$. The single-electron density of states at these
energies goes linearly with $\omega$: $\rho(\omega) \sim |\omega|$,
giving rise to extra $T$-factors in thermodynamic quantities
of pure d-wave superconductors.

 Let us introduce the Nambu spinor
\[
\Phi_{\bf k} = \left(
\begin{array}{c}
c_{{\bf k}\uparrow}\\
c^+_{-{\bf k}\downarrow}
\end{array}
\right).
\]
The Hamiltonian (1) in this notations is given by
\begin{equation}
H = \sum_{\bf k}\Phi_{\bf k}^+[(\epsilon_{\bf k} - \mu)\tau_3 +
\Delta({\bf k})\tau_2]\Phi_{\bf k},
\end{equation}
where $\tau_i$ are the Pauli matrices. We linearize the spectrum close to
the nodes and pass to a continuum description in terms of four Fermi fields
$\psi_j(x),$ $j=1,\bar 1, 2, \bar 2$. We
choose a new coordinate system rotated by angle $\pi/4$
with respect to the original one. It is also convenient
to make a
$\pi/2$ rotation in the Nambu space about the $\tau_2-$axis.
The resulting continuum
model is then given by the sum of four Dirac-like
Hamiltonians:

\begin{eqnarray}
H  &=&
\sum_{j=1,\bar 1, 2, \bar 2}\int d^2x \psi^+_j(x) \hat H_j(x) \psi_j(x), \\
\hat H_1(x) &=& - \hat H_{\bar 1}(x) =
- iv_1\partial_1\tau_1 - iv_2\partial_2\tau_2, \\
\hat H_2(x) &=& - \hat H_{\bar 2}(x) =
- iv_1\partial_2\tau_1 - iv_2\partial_1\tau_2,
\end{eqnarray}
with $v_1 = 2\sqrt 2 ta\sin k_0, v_2 = \sqrt 2\Delta_0\sin k_0$.

Processes of scattering on nonmagnetic  impurities are described by
\begin{equation}
H_{imp} = \sum_{jj'}\int d^2x V_{jj'}(x)\psi^+_j(x)\tau_1\psi_{j'}(x),
\end{equation}
where $V_{jj'}(x) = V_{j'j}^*(x)$ are random fields with  Gaussian
distributions. Since $V_{jj'}(x)$ are wave packets with average wave
vectors ${\bf k}_j - {\bf k}_{j'}$, not all of them are independent. For the
square geometry presented on Fig. \ref{figure1}, we
have the following independent
components:

(i) the real field $V_{jj}(x) \equiv V_0(x)$ representing a slowly varying
(quasiclassical) component of the impurity potential;

(ii) the complex fields $V_{1\bar 1}(x) \equiv V_1(x)$ and
$V_{2\bar 2}(x) \equiv V_2(x)$ representing a back scattering with
the momentum transfer $2k_0$;

(iii)  the complex fields $V_{12}(x) = V_{\bar 2\bar 1} \equiv W_1(x)$ and
$V_{1\bar 2}(x) = V_{2\bar 1}(x) \equiv W_2(x)$ corresponding to
transitions between nearest nodes.

In what follows we assume that all these fields are distributed according
to the Gauss' law:
\begin{eqnarray}
<V_0(x)V_0(x')> &=& \lambda_0\delta^{(2)}(x - x'),\nonumber\\
<V_j(x)V_j^*(x')> &=& \lambda_1\delta^{(2)}(x - x'),~ (j = 1,2),\nonumber\\
<W_j(x)W_j^*(x')> &=& \lambda_2\delta^{(2)}(x - x'),~ (j = 1,2).
\end{eqnarray}

\section{Self Energy Diagrams}

For a given realization of static random potential, the one-particle
Green's function tensor written
in the mixed $x-\omega_n$ representation
($\omega_{n}=(2n+1)\pi/\beta$ being the
Matsubara frequency) satisfies the following integral equation
\begin{eqnarray}
\hat{G}_{jj'}(x,x';\omega_{n}) &=&
\delta_{jj'}~ \hat{G}_{j}^{0}(x - x';\omega_{n})
\nonumber\\
& &+ \sum_{j''} \int d^{2}x''~ \hat{G}_{j}^{0}(x - x'';\omega_{n})
V_{jj''}(x'') ~\hat{\tau}_{1}~\hat{G}_{j''j'}(x'',x';\omega_{n}),
\label{dyson}
\end{eqnarray}
where
\begin{equation}
\hat{G}_{j}^{0}(x - x';\omega_{n})
= [i\omega_{n} - \hat{H}_{j}(x)]^{-1} \delta (x - x')
\end{equation}
is the Green's function for the pure system.
\par
For weak electron-impurity scattering, the standard procedure
 (see, e.g. \onlinecite{AGD}) consists in
developing a perturbation theory expansion in Eq.(\ref{dyson}), with subsequent
averaging of
products of the Gaussian random fields $V_{jj'}(x)$, appearing in
each term of this series.
The averaged Green's function
\begin{equation}
<\hat{G}_{jj'}(x,x';\omega_{n})> = \delta_{jj'} \hat{G}_{j}(x - x';\omega_{n}),
\end{equation}
written in momentum representation, satisfies the Dyson equation
\begin{equation}
\hat{G}_{j}^{-1} ({\bf k}, \omega_{n}) =
i\omega_{n} - \hat{H}_{j}({\bf k})
- \hat{\Sigma}_{j}({\bf k}, \omega_{n}).
\end{equation}
Here
\begin{equation}
\hat{H}_{1}({\bf k}) = - \hat{H}_{\bar {1}}({\bf k})
= k_{1}v_{1}\hat{\tau}_{1} + k_{2}v_{2}\hat{\tau}_{2},
\end{equation}
\begin{equation}
\hat{H}_{2}({\bf k}) = - \hat{H}_{\bar {2}}({\bf k})
= k_{2}v_{1}\hat{\tau}_{1} + k_{1}v_{2}\hat{\tau}_{2}.
\end{equation}
$\hat{\Sigma}_{j}({\bf k}, \omega_{n})$ is the self-energy operator
incorporating all effects
of the electron-impurity scattering.
In what follows, we shall analyze  lowest-order diagrams for
$\hat{\Sigma}_{1}$. In calculations, a cutoff
prescription, defined by conditions $|H_j({\bf k})| \leq \Delta_0$,
will be used.
\par
In the lowest (Born) approximation $\hat{\Sigma}_{1}$
is given by the first diagram
shown in Fig. \ref{figure2}. It is momentum independent
and contains a logarithmic
singularity, as previously pointed out in a
number of papers \cite{Gorkov85,Ueda85,Gross86,Kusma85,Lee93}:
\begin{eqnarray}
\hat{\Sigma}_{1}^{(1)}({\bf k}, \omega_{n}) &=&
\hat{\Sigma}_{1}^{(1)}(\omega_{n})\nonumber\\
&=& \int \frac{d^{2}k'}{(2\pi)^{2}} \hat{\tau}_{1}
[\lambda_{0}\hat{G}_{1}^{0} ({\bf k'}, \omega_{n})
+ \lambda_{1}\hat{G}_{\bar{1}}^{0} ({\bf k'}, \omega_{n}) \nonumber\\
& &\; + \lambda_{2} (\hat{G}_{2}^{0} ({\bf k'}, \omega_{n}) +
\hat{G}_{\bar{2}}^{0} ({\bf k'}, \omega_{n}))]\hat{\tau}_{1}\nonumber\\
&=& -i\omega_{n}~ (g_{3} + g_{1} + 2g_{2}) \int_{|p| < \Delta_{0}}
\frac {d^{2}p}{(2\pi)^{2}} \frac{1}{p^{2} + \omega^{2}_{n}}\nonumber\\
&=& -i\omega_{n}~ \frac{g_{3} + g_{1} + 2g_{2}}{2\pi}~
\ln~\frac{\Delta_{0}}{|\omega_{n}|}. \label{eq:diag1}
\end{eqnarray}
Here
\begin{equation}
g_{1,2} = \frac{\lambda_{1,2}}{v_{1}v_{2}}, ~~
 g_{3} = \frac{\lambda_{0}}{v_{1}v_{2}}
\end{equation}
are small dimensionless coupling constants, proportional to the
impurity concentration $c$.

In fact,
logarithmic singularities appear in all orders of perturbation theory. Consider
second-order diagrams shown in Fig. \ref{figure2}.
The "rainbow" diagram is obtained from the Born
one by first-order renormalization of the internal electron line.
The node indices $j_{1}$ and $j_{2}$ take arbitrary values from the set
$(1, \bar{1}, 2, \bar{2})$. The result is easily found to be
\begin{equation}
\Sigma^{(2a)}(\omega_{n}) =
-i\omega_{n}~ \frac{(g_{3} + g_{1} + 2g_{2})^{2}}{(2\pi)^{2}}~
[~\ln^{2}\frac{\Delta_{0}}{|\omega_{n}|} + 0(1)~]. \label{eq:diag2}
\end{equation}
\par
The diagram with crossing impurity lines, shown in  Fig. \ref{figure2},
corresponds to a vertex renormalization in the Born
self-energy diagram. The momentum conservation restricts the
allowed distribution
of the node indices $(j_{1},j_{2},j_{3})$ by the following nine possibilities:
$(111), (\bar{1} \bar{1} 1), (1 \bar{1} \bar{1}), (221),$ $ (122),
(\bar{2} \bar{2} 1), (1 \bar{2} \bar{2}), (2 \bar{1} \bar{2}),
(\bar{2} \bar{1} 2)$.
Using power counting and symmetry arguments, one can show
that, in each diagram, the maximum power of $\log |\omega_{n}|$
appears with the prefactor $\sim i\omega_{n}$. This allows to simplify
calculations by setting the external
momentum ${\bf k} = 0$, in which case
\begin{eqnarray}
\hat{\Sigma}^{(2b)} ({\bf k} = 0, \omega_{n}) \mid_{(j_{1},j_{2},j_{3})}
&=& \int \frac{d^{2}k}{(2\pi)^{2}} \frac{d^{2}q}{(2\pi)^{2}}
\nonumber\\ & &\times
\hat{\tau}_{1}~ \hat{G}^{0}_{j_{1}}({\bf k},\omega_{n})~
\hat{\tau}_{1}~ \hat{G}^{0}_{j_{2}}({\bf q}, \omega_{n})~
\hat{\tau}_{1}~ \hat{G}^{0}_{j_{3}}({\bf q} -
{\bf k},\omega_{n})~ \hat{\tau}_{1}.
\end{eqnarray}
Consider, for example, the impurity scattering in the vicinity of
a single node, corresponding
to the case (111). Here we are able to rescale all the momenta
to get a circular cutoff in each integral:
\begin{eqnarray}
\Sigma^{(2b)}_{(111)}(\omega_{n}) = -i\omega_{n} g^{2}_{3}
\int_{(|k| <\Delta_{0})} \frac{d^{2}k}{(2\pi)^{2}}
\int_{(|q| <\Delta_{0})} \frac{d^{2}q}{(2\pi)^{2}}\nonumber\\
\times \frac{{\bf k} \cdot ({\bf q} - {\bf k})}
{(\omega^{2}_{n} + k^{2})[\omega^{2}_{n} + ({\bf q} - {\bf k})^{2}]
(\omega^{2}_{n} + q^{2})}.
\end{eqnarray}
Integrating over the angle between ${\bf k}$ and ${\bf q}$ yields
$-2\pi \Theta_{\omega_{n}}(k^{2} - q^{2})$, where $\Theta_{\omega}(x)$
is a smeared
step function, with the width of the order $|\omega|$ at $x = 0$. The remaining
integration over the region $|{\bf k}| > |{\bf q}|$ gives a
$\log^{2}|\omega_{n}|$
contribution:
\begin{eqnarray}
\Sigma^{(2b)}_{(111)}(\omega_{n}) &\simeq& i\omega_{n}
{}~\frac{g_{3}^{2}}{(2\pi)^{2}}
\int_{0}^{\Delta_{0}} \frac{q~dq}{\omega_{n}^{2} + q^{2}}
\int_{q}^{\Delta_{0}}
\frac{k~dk}{\omega_{n}^{2} + k^{2}}
\nonumber\\
&=& ~i\omega_{n} \frac{g_{3}^{2}}{2(2\pi)^{2}}~
\ln^{2} \frac{\Delta_{0}}{|\omega_{n}|}. \label{eq:diag3}
\end{eqnarray}
Quite similarly
\begin{equation}
\Sigma^{(2b)}_{(\bar{1} \bar{1} 1)}(\omega_{n}) +
\Sigma^{(2b)}_{(1 \bar{1} \bar{1})}(\omega_{n})
\simeq -i\omega_{n}~ \frac{g_{3}g_{1}}{(2\pi)^{2}}~
\ln^{2} \frac{\Delta_{0}}{|\omega_{n}|}.
\end{equation}
\par
In cases when intermediate states belong to neighboring nodes, as it
takes place for
diagrams $(221)$, $(122)$,
$(\bar{2} \bar{2} 1)$,$ (1 \bar{2} \bar{2})$,$ (2 \bar{1} \bar{2})$,$
(\bar{2} \bar{1} 2)$
the situation is somewhat more complicated due to the difference
between the two velocities,
$v_1 \gg v_2$. The two ellipses, representing the surfaces
of constant energy for states $1~(\bar{1})$ and $2~(\bar{2})$,
are strongly elongated in
perpendicular directions and, therefore, cannot be simultaneously transformed
to circles by a global rescaling of the momenta. The available phase
space is then reduced
by the amount determined by the anisotropy parameter
\begin{equation}
\gamma = v_{2}/v_{1} \sim \Delta_{0}/t \ll 1.
\end{equation}
As a result, an additional energy scale $\gamma\Delta_{0}$ enters
the problem, and
the corresponding diagrams become dependent on the ratio
$|\omega_n|/\gamma\Delta_{0}$.
For example,
\begin{eqnarray}
\Sigma^{(2b)}_{(2 \bar{1} \bar{2})}(\omega_{n}) +
\Sigma^{(2b)}_{(\bar{2} \bar{1} 2)}(\omega_{n})
&=& 2i\omega_{n}~g^{2}_{2}~
\int_{(|k|<\Delta_{0})} \frac{d^{2}k}{(2\pi)^{2}}
\int_{(|q_{1}|<\gamma\Delta_{0},|q_{2}|<\Delta_{0})}
\frac{d^{2}q}{(2\pi)^{2}}\nonumber\\
& &\times\frac{{\bf k} \cdot ({\bf q} - {\bf k})}
{(\omega_{n}^{2} + k^{2})
(\omega_{n}^{2} + \gamma^{-2} q^{2}_{1} + \gamma^{2} q^{2}_{2})
[\omega_{n}^{2} +({\bf q} - {\bf k})^{2}]}.
\end{eqnarray}
Integrating first over ${\bf k}$, we get
\begin{equation}
\Sigma^{(2b)}_{(2 \bar{1} \bar{2})}(\omega_{n}) +
\Sigma^{(2b)}_{(\bar{2} \bar{1} 2)}(\omega_{n})
\simeq - 4~\frac{2i\omega_{n}}{(2\pi)^{3}}~g^{2}_{2}
\int_{0}^{\gamma\Delta_{0}} dq_{1} \int_{0}^{\Delta_{0}} dq_{2}~
\frac{\ln~[\Delta_{0}/max(|\omega_{n}|,|q|)]}{\omega_{n}^{2}
+\gamma^{-2} q^{2}_{1} + \gamma^{2} q^{2}_{2}}.\label{above}
\end{equation}
At energies $ \gamma \Delta_0 < |\omega_n| \ll \Delta_0 $,
the integral in \ref{above} is mostly contributed
by regions $q_1 \ll \gamma \Delta_0,~
\gamma \Delta_0 \ll q_2 \ll \Delta_0$, and is nearly independent of $\omega_n$:
\begin{equation}
\Sigma^{(2b)}_{(2 \bar{1} \bar{2})}(\omega_{n}) +
\Sigma^{(2b)}_{(\bar{2} \bar{1} 2)}(\omega_{n}) \simeq
- \frac{2ig^{2}_2}{(2\pi)^{2}}~ \gamma \Delta_0~ [1 + 0(\gamma \ln \gamma)]~
\mbox{sign}(\omega_{n}).
\end{equation}
The $\log^{2}$-term only appears at lower energies, $|\omega_{n}| \ll \gamma
\Delta_{0}$,
in which case integration over $|q| < \gamma \Delta_0$ yields
\begin{equation}
\Sigma^{(2b)}_{(2 \bar{1} \bar{2})}(\omega_{n}) +
\Sigma^{(2b)}_{(\bar{2} \bar{1} 2)}(\omega_{n}) \simeq
- i\omega_{n}~ \frac{g^{2}_{2}}{(2\pi)^{2}}~
\ln^{2} \frac{\gamma \Delta_{0}}{|\omega_{n}|}.
\end{equation}
For the remaining four diagrams ($(221),(122), (\bar{2} \bar{2} 1),
(1 \bar{2} \bar{2})$)
the effect of the anisotropy $\gamma$ is even stronger. It can be shown that
at $|\omega_{n}| \ll \gamma \Delta_{0}$ these diagrams contain
only first power of the logarithm: $\Sigma \sim i\omega_{n} g_{3} g_{2} \ln
(\gamma\Delta_{0}/|\omega_{n}|)$.

\par
The above analysis shows that, in 2D electron systems with
a conical spectrum, impurity scattering drastically differs from that in a
normal
metallic state with a finite Fermi surface.
The structure of the perturbation theory expansion for
the self energy reveals a "logarithmic situation" resembling mass
renormalization
in models of interacting fermions in one space dimension. (This analogy,
being realized in earlier papers \cite{Fisher85,Fradkin86},
will become more transparent in subsequent sections, where the effective field
theory in (1 + 1) dimensions is discussed).
Notice that,
after analytic continuation, $i\omega_{n} \rightarrow \omega + i\delta$,
the leading logarithmic singularities occur in the real part of
$\Sigma (\omega)$. This means that, in systems with a conical spectrum,
disorder mostly affects renormalization of the single-particle spectrum,
as opposed to the usual picture in systems with a finite DOS,
where life-time effects dominate.
\par
Actually, the logarithmic singularities appear in all orders of
perturbation theory,
including crossing diagrams. In two- and higher-dimensional
metallic systems with a finite
Fermi surface, such diagrams are known to be relatively small
\cite{AGD}. In the case of weak
disorder,  $k_{F} l \gg 1$, $l$ being the mean-free path,
strong restrictions over
the relative angles between the momenta of intermediate one-electron states
effectively reduce the
available phase space, making diagrams with crossing impurity
lines proportional to
powers of the factor $(1/k_{F}l)$.
However, in 1D systems, where the Fermi surface is represented
by two points, $\pm k_{F}$,
no such reduction is possible, and all diagrams are equally important,
as is known
from the theory of one-dimensional localization
\cite{Berezinskii,Abrikosov}.
In this respect, the electron-impurity scattering in
2D systems with conical spectra resembles the 1D case: the crossing diagrams
also contain logarithmic singularities and therefore should be treated on
the same footing
as the rainbow ones. For a finite number of degeneracy points (nodes), no
selection
of diagrams is possible. An approximation, in which one confines
consideration to the class of
rainbow diagrams and then solves self-consistently the resulting Dyson equation
for $\Sigma (\omega)$ \cite{Lee93}, can be justified only in the limit of
(infinitely)
large number of nodes \cite{Fradkin86},
or when one extends consideration to a higher dimensional case, {\it e.g.}
 3D polar superconductors with a $line$ of nodes \cite{Gorkov85}.
This approximation leads to a finite DOS at $\omega = 0$.
\par
In the next sections, all singular self-energy diagrams will be taken into
account,
using replica trick. Treating the effective, disorder-free field-theoretical
model by abelian and non-abelian bosonization, we shall show that,
for a weakly disordered 2D d-wave superconductor, the low-energy DOS
follows a power-law behavior. The corresponding critical exponent
depends on the
type of electron-impurity processes included into consideration.

\section{Replica Trick and Effective Two-Dimensional Field Theory}

In what follows, we shall be mostly concerned with the forward and
 backward scattering processes,
putting $\lambda_2 = 0$. In this case the nodes $(1, \bar 1)$ and
$(2, \bar 2)$ can
be considered independently. Without loss of generality we can choose the
 $(1, \bar 1)$ pair. Rescaling the coordinates and the fields
$$
x \rightarrow v_1x_1,~~ y \rightarrow v_2x_2,
$$
$$
 \psi_1(x) \rightarrow \frac{1}{\sqrt{v_1v_2}}\psi_1(x),~~
\psi_{\bar 1}(x) \rightarrow \frac{1}{\sqrt{v_1v_2}}
\psi_{\bar 1}(x)
$$
we arrive at following  generating functional for the (unaveraged) Green's
function with a fixed Matsubara frequency $\omega_n$:
\begin{eqnarray}
Z[A] &=& \int D\bar\eta D\eta\exp( - S[A]),\\
S &=&
\int d^2x \bar\eta(x)[(- i\partial_1 + A_3(x)\sigma_3)\tau_1\nonumber\\
&&+ ( - i\partial_2 + A_1(x)\sigma_1 +  A_2(x)\sigma_2)\tau_2 -
i\omega_n]\eta(x). \label{eq:action}
\end{eqnarray}
Action (\ref{eq:action}) describes
two-dimensional massless fermions interacting
with a random non-Abelian gauge potential.
The Pauli matrices $\sigma_i$ act on the isotopical indices
$(1, \bar 1)$, and the probability distribution of the gauge field is
\begin{equation}
P[A] = \int DA \; \exp[ - \int d^2 x\frac{A_i(x)^2}{2g_i}].
\end{equation}
The matrices $\tau_2 = \gamma_0$ and $\tau_1 = \gamma_1$ form a
representation of the
two dimensional Clifford algebra, i.e they are
the Dirac matrices of our problem. The new  fermionic fields are related
to the initial ones as follows:
\begin{equation}
\bar\eta = - i\psi^+,~~\eta = \psi.
\end{equation}
In terms of the new fermionic variables, the DOS is given by
\begin{equation}
\rho(\omega) = - \frac{1}{\pi v_1v_2}Re[Tr<\bar\eta(x)\eta(x)>]|_{i\omega_n
\rightarrow  \omega + i\delta}. \label{eq:rho}
\end{equation}

We average over the disorder using the standard replica trick. For this
purpose  we replicate the action (\ref{eq:action}) $r$-times
($r$ is the number of replicas)
and integrate the fermionic partition function with the
replicated action over the gauge fields. The result is
\begin{equation}
Z_r = \int D\bar\eta D\eta\exp( - S_r),
\end{equation}
\begin{eqnarray}
S_r &=&
\int d^2x \{\bar\eta_{a,\alpha}
(\gamma_{\mu}\partial_{\mu} + \omega_n)\eta_{a,\alpha} +
\frac{g_3}{2}(\bar\eta_{a,\alpha}\gamma_{0}\sigma^3_{\alpha\beta}
\eta_{a,\beta})^2  \nonumber\\
& &+ \frac{g_1}{2} [(\bar\eta_{a,\alpha}\gamma_1\sigma^1_{\alpha\beta}
\eta_{a,\beta})^2 +
(\bar\eta_{a,\alpha}\gamma_1\sigma^2_{\alpha\beta_{a,\beta})^2}
\}, \label{eq:Ac-rep}
\end{eqnarray}
where the Greek indices are isotopic and the Latin ones are reserved for
replicas ($a = 1, ... r$). It will be convenient to rewrite  the
problem in the Hamiltonian formalism treating $x_1$ asimaginary time
and $x_2$ as space coordinate. The quantization rules are
$
\bar\eta\gamma_0 = \eta^+,
$
where the spinors $(\eta_R, \eta_L)$ and their Hermitian conjugate
satisfy the standard anticommutation relations. The Hamiltonian
is given by
\begin{eqnarray}
H &=& \int dx \{\eta^+_{a,\alpha}(- i\partial_x\tau_3 + \omega_n\tau_2)
\eta_{a,\alpha}
\nonumber\\
& &\; - \frac{g_3}{2}(:\eta_{a,\alpha}^+\tau_3\sigma^3_{\alpha\beta}
\eta_{a,\beta}:)^2 + \frac{g_1}{2}~[ (:\eta_{a,\alpha}^+\sigma^1_{\alpha\beta}
\eta_{a,\beta}:)^2 + \frac{g_1}{2}(:\eta_{a,\alpha}^+\sigma^2_{\alpha\beta}
\eta_{a,\beta}:)^2]~\},\label{eq:Ham1D}
\end{eqnarray}
where the Matsubara frequency $\omega_n$ plays the role of relativistic
mass.

Notice that the Hamiltonian (\ref{eq:Ham1D}) possesses continuous $SU(r)$
replica symmetry.
At $\omega_n = 0$ it is also invariant under chiral rotations of Fermi fields:
(the continuous $\gamma^5-$symmetry) $\eta \rightarrow \exp (i \tau_3 \varphi)~
\eta$. As follows from (\ref{eq:rho}), a
finite $\rho(0)$ would mean the existence of a nonzero order parameter
$<\bar{\eta} \eta>$, indicating a spontaneous breakdown of
these symmetries. However, continuous symmetries cannot be
broken in two dimensions; therefore $\rho(\omega)$ should vanish at
$\omega \rightarrow0$. This is confirmed by direct calculations
presented below.

\section{Model without Backscattering. Bosonization}

 In this Section we shall focus on the slowly varying (quasiclassical)
part of the random
potential, i.e. we set $g_1 = g_2 = 0$. In this case the problem is reduced to
the single node one. The isotopic indices can be suppressed, and the resulting
model can be treated by the standard abelian bosonization methods .

 The bosonization rules are well known\cite{Frishman93}:

\begin{eqnarray}
J_{R,a} + J_{L,a} = \frac{1}{\sqrt\pi}~\partial_x\phi_a &,&
J_{R,a} - J_{L,a} = - \frac{1}{\sqrt\pi}~\Pi_a ,\nonumber\\
:\eta_{R,a}^+\eta_{L,a} + \eta_{L,a}^+\eta_{R,a}: &=&
- \Lambda\cos(\sqrt{4\pi}\phi_a). \label{eq:rules}
\end{eqnarray}
Here
\begin{equation}
J_{R(L),a} = :\eta_{R(L),a}^+\eta_{R(L),a}:
\end{equation}
are the current operators for right(left)-moving particles with the
replica index $a$;
$\phi_a(x)$ and $\Pi_a(x)$ are scalar fields and their conjugate momenta,
respectively,
satisfying the canonical commutation relations:
$
[\phi_a(x), \Pi_b(y)] = i\delta_{ab}\delta(x - y);
$
$\Lambda \sim \Delta_{0}$ is the ultraviolet cut-off.
The Bose model equivalent to (\ref{eq:Ham1D}) reads:
\begin{eqnarray}
H = \int dx\{ \frac{1}{2}\sum_a[\Pi_a^2 + (\partial_x\phi_a)^2] -
\frac{g_3}{2\pi}\sum_{ab}\Pi_a\Pi_b
- \omega_n\Lambda\sum_a\cos(\sqrt{4\pi}\phi_a)\}.
\end{eqnarray}

In order to proceed further we need to diagonalize the quadratic form
$\sum_{ab}\Pi_a\Pi_b \equiv \Pi^TM\Pi$, where $M_{ab} = 1$ for all $a$ and $b$.
 We do it with an orthogonal transformation $U$:
\begin{eqnarray}
\Pi &=& UP,~~ U^TU = I; \nonumber\\
U^TMU &=& M_D, ~~(M_D)_{ab} = r\delta_{ar}\delta_{br}; \nonumber\\
U_{ar} &=& \frac{1}{\sqrt r}~~ (a = 1, ... r).
\end{eqnarray}
and obtain
\begin{eqnarray}
H = \int dx & &\{ \frac{1}{2}\sum_{a = 1}^{r - 1}~
[P_a^2 + (\partial_x\phi_a)^2] +
 \frac{1}{2}~[(1 - \frac{rg_3}{\pi})P_r^2 + (\partial_x\phi_r)^2]
\nonumber\\
& &- \omega_n\Lambda\sum_a\cos(\sqrt{4\pi}U_{ab}\phi_b)\}.
\end{eqnarray}
The last step consists in rescaling  $\phi_r$ and $P_r$:
\begin{eqnarray}
\phi_r \rightarrow  \gamma\phi_r &,& ~~P_r \rightarrow \frac{1}{\gamma}P_r,
\nonumber\\
\gamma^2 &=& (1 - \frac{g_3r}{\pi})^{1/2}.
\end{eqnarray}
The resulting theory is described by the Hamiltonian
\begin{equation}
H =
\int dx\{\sum_{a = 1}^{r}\frac{u_a}{2}[P_a^2 + (\partial_x\phi_a)^2] -
\omega_n\Lambda\sum_a\cos(\sqrt{4\pi}V_{ab}\phi_b)\}, \label{eq:Bose}
\end{equation}
where
$$
u_a = 1~ ( a < r),~~ u_r = \gamma,
$$
$$
V_{ab} = U_{ab}~ (a, b < r),~~ V_{ar} = \frac{\gamma}{r}.
$$
The critical dimension of the cosine term in Eq.(\ref{eq:Bose}) is given by
\begin{equation}
\Delta_a  = \sum_{b = 1}^r V^2_{ab} = 1 + \frac{\gamma^2 - 1}{r}.
\end{equation}
We see that it does not depend on $a$ and remains well defined in
the replica limit:
\begin{equation}
\Delta \equiv \Delta_a(r \rightarrow 0) = 1 - \frac{g_3}{2\pi}.
\end{equation}

 Simple scaling arguments allow us to estimate the DOS at finite $\omega$.
First we note that in Eq.(\ref{eq:Bose}) the cosine term is a relevant
perturbation generating
a new energy scale $\tilde{\omega}$
(or, equivalently, correlation length $\xi_c$ ), given by
\begin{equation}
\tilde{\omega} \sim \xi_c^{-1}
\sim \omega \left(\frac{\Lambda}{|\omega|} \right)
^{\frac{1 - \Delta }{2 - \Delta}}. \label{eq:length}
\end{equation}
This formula correctly reproduces perturbation theory expansion for
the self-energy
operator in the single-node approximation ($g_1 = g_2 = 0$):
\begin{eqnarray}
\frac{\Sigma (i\omega_n)}{i\omega_n}
\mid_{i\omega_n \rightarrow \omega + i\delta}~ &=&
1 - \frac{\tilde{\omega}}{\omega}
= 1 - \left(\frac{\Lambda}{|\omega|} \right)
^{\frac{1 - \Delta }{2 - \Delta}}\nonumber\\
&=& - \frac{g_3}{2\pi} \ln \frac {\Lambda}{|\omega|} -
\frac{1}{2} \left(\frac {g_3}{2\pi} \right)^2
\ln^2 \frac {\Lambda}{|\omega|} + \cdots ,
\end{eqnarray}
in agreement with the estimation of first- and second-order diagrams (see
 Eqs.(\ref{eq:diag1}),(\ref{eq:diag2}) and (\ref{eq:diag3})).
Applying then bosonization rules (\ref{eq:rules}) to  Eq.(\ref{eq:rho})
we establish
that $\rho(\omega) \sim \left< \cos(\sqrt{4\pi}V_{ab}\phi_b) \right>$.
This average
is proportial to $\xi^{- \Delta}$. Using then (\ref{eq:length}),
we find a power-law behavior
of the low-energy DOS
\begin{equation}
\rho(\omega) \sim |\omega|^{\alpha}
\end{equation}
with a nonuniversal exponent
\begin{equation}
\alpha = \frac{\Delta}{2 - \Delta} = \frac{1 - g_3/2\pi}{1 + g_3/2\pi} < 1,
\end{equation}
depending on the impurity concentration.

\section{Impurity Scattering in Model with Several Fermi Points}

In this Section we consider the model
where scattering processes transfer electrons
between the opposite Fermi points. The corresponding replicated  model
was described  in
Section IV, Eq. (\ref{eq:Ham1D}); here we rewrite it explicitly
in terms of chiral fermions:
\begin{eqnarray}
H &=& H_0 + H_1 + H_2,\label{eq:Ham2} \\
H_0  &=& \int dx \eta^+_{a,\alpha}(- i\partial_x\tau_3 + \omega_n\tau_2)
\eta_{a,\alpha}, \label{eq:HamFree} \\
H_1 &=&
\int dx \{ - 2 g_3 (:J_R^3J_R^3: + :J_L^3J_L^3:) +  2 g_1 \sum_{i = 1,2}
(:J_R^iJ_R^i: + :J_L^iJ_L^i:)\}, \label{eq:int1}\\
H_2 &=& 4 \int dx [ g_3 :J_R^3J_L^3: +  g_1\sum_{i = 1,2}
:J_R^iJ_L^i:], \label{eq:int2}
\end{eqnarray}
where
\begin{eqnarray}
J_R^i &=& \eta_{R,a\alpha}^+
\frac{\sigma^i_{\alpha\beta}}{2}\eta_{R,a\beta},\nonumber\\
J_L^i &=& \eta_{L,a\alpha}^+\frac{\sigma^i_{\alpha\beta}}{2}\eta_{L,a\beta}
\label{eq:currents}
\end{eqnarray}
are chiral components of the $SU(2)$ currents.

 We shall apply to the model (\ref{eq:Ham2}) the procedure of non-Abelian
bosonization developed by Witten \cite{Witten}. The approach is based on the
following key moments. The first one is that the Hamiltonian of free
fermions (\ref{eq:HamFree}) can be expressed in terms of current operators
(the Sugawara construction \cite{Frishman93,Affleck861}).
The Sugawara
form of the Hamiltonian of
free massless fermions with a general $U(1)\times SU(N)
\times SU(r)$-symmetry (in Eq.(\ref{eq:HamFree}) N = 2) is

\begin{eqnarray}
H_0  &=&
 H^0_{U(1)} + H^0_{SU(N)} + H^0_{SU(r)}\nonumber\\
&=& \int dx[ \frac{\pi}{Nr}
(:J_R(x)J_R(x): + :J_L(x)J_L(x):) \nonumber\\
& &+ \frac{2\pi}{N + r}\sum_{i = 1}^{G_N}( :J^i_R(x)J^i_R(x):
+ :J^i_L(x)J^i_L(x):)
\nonumber\\
& &+ \frac{2\pi}{N + r}\sum_{a = 1}^{G_r}( :J^a_R(x)J^a_R(x):
+ :J^a_L(x)J^a_L(x):)],
\label{eq:HamSug}
\end{eqnarray}
where  $J$, $J^i$ and $J^a$ are the $U(1)$, the $SU(N)$ and the $SU(r)$
currents defined as in Eq.(\ref{eq:currents}), but with  generators of
the corresponding algebras instead of the Pauli matrices. $G_N$ and $G_r$ are
the total number of generators of the $su(N)$ and the $su(r)$ Lie algebras.
The second key
moment is that the currents from different algebras commute. Therefore $H_0$
is a sum of three mutually commuting operators.

 Now notice that the interaction terms (\ref{eq:int1}),(\ref{eq:int2})
contain only spin currents and
therefore do not affect the spectra of $SU(2)$-singlets.
Since this interaction is attractive,
 the spectrum in the
$SU(2)$ channel becomes gapful and, as we show later,
 this gap persists in the replica limit.
Therefore, if $\omega_n = 0$,
the spectrum well below the gap $Q_0$ is described by the rest of
the Hamiltonian (\ref{eq:Ham2}), in other words by
$
H_{eff} =  H^0 _{U(1)} + H^0 _{SU(r)}.
$
The Hamiltonian $ H^0 _{SU(r)}$ is the Hamiltonian of the Wess-Zumino-Witten
model.
Its spectrum is the subsector of the free fermionic spectrum generated
by the $SU(r)$ current operators.  The model is conformally invariant and
exactly solvable \cite{Tsv87}. Its primary fields transform according to
representations of the $SU(r)$
group. In particular, the field from the
fundamental representation, which we denote $g_{ab}$, has the following
scaling dimension \cite{Knizh84}

\begin{equation}
\Delta_g = \frac{r - 1/r}{N + r}.
\end{equation}
We see that $\Delta_g$ is singular at $r \rightarrow 0$, which indicates
that the $g$-field cannot appear alone in the physical sector.
In fact, as we shall see, a local bilinear form of two Fermi operators,
$\bar{\eta} \eta$,
which appears in the
"mass" (i.e. $\omega_n$-proportional) term of the Hamiltonian, as well as in
the definition (30) of DOS, is represented by a combination of $g$ and a phase
exponential of the scalar ($U(1)$-symmetric) field $\phi$
\begin{equation}
\bar{\eta} \eta \equiv Q \sim Q_0~ Re~ [ g \exp (i \sqrt{\frac{4\pi}{Nr}}
\phi)].
\label{eq:orderpar}
\end{equation}
The scaling dimension of the exponential is equal to $1/Nr$.
So, the two  singularities cancel, and
the resulting scaling dimension of the operator  $Q$
 is finite in the replica limit:
\begin{equation}
\Delta_Q = \lim_{r \rightarrow 0}~[\frac{1}{Nr} +  \frac{r - 1/r}{N + r}] =
\frac{1}{N^2}.
\label{eq:dim}
\end{equation}
\par
 It is instructive to repeat the above derivation in a more
traditional form introducing the replica tensor $Q_{ab}$. For the sake of
simplicity we assume $g_1 = g_3$.  Let us
rewrite the action (\ref{eq:Ac-rep}) in the form suitable for the
Hubbard-Stratonovich transformation:
\begin{eqnarray}
S &=&
\int d^2x \{\bar\eta_{a,\alpha}(i\gamma_{\mu}\partial_{\mu} + i\omega_n)
\eta_{a,\alpha}
\nonumber\\
& &\; -
g[(\bar\eta_{a,\alpha}\eta_{b,\alpha})(\bar\eta_{b,\beta}\eta_{a,\beta}) -
(\bar\eta_{a,\alpha}\gamma_5\eta_{b,\alpha})(\bar\eta_{b,\beta}\gamma_5
\eta_{a,\beta})] + S_{irr},\label{eq:nonabel}
\end{eqnarray}
where $\gamma_5 = i\tau_1$ and $S_{irr}$ contains terms with interactions
near the same Fermi point. Such terms renormalize velocities and are not
important for the current discussion. Now we introduce the auxilary field
 $Q_{\alpha\beta}$ and  decouple the interaction
term by the Hubbard-Stratonovich transformation. Integrating formally over
the fermions, we obtain the partition function for the replica tensor field:
\begin{eqnarray}
Z &=& \int DQ^+DQ\exp( - \int d^2x L),\nonumber\\
L &=&  \frac{1}{2g}Tr(Q^+ - \omega I)(Q - \omega I)  \nonumber\\
& &-N \; Tr\ln[i\gamma_{\mu}\partial_{\mu} +
(1 + i\gamma_5)Q/2 + (1 - i\gamma_5)Q^+/2]. \label{eq:z}
\end{eqnarray}
Following the conventional wisdom, we shall look for a
saddle point of the exponent and then  derive an effective action for
fluctuations
around
this saddle point. We assume that  the saddle point configuration of $Q$ is
coordinate independent and
as such can be chosen as a diagonal real matrix:
$Q(x) = diag(\lambda_1, ... \lambda_r)$. The
density of effective action on this configuration is equal to
\begin{equation}
S_{eff} = \sum_{a}[\frac{(\lambda_{a} - \omega)^2}{2g} +
\frac{N\lambda^2_{a}}{2\pi}\ln\frac{|\lambda_{a}|}{\Lambda}].
\end{equation}
The saddle point value of $Q$, being a point of minimum of this function,
satisfies the equation:
\begin{equation}
\frac{\lambda_{a} - \omega}{g} + \frac{N\lambda_{a}}{\pi}
\ln\frac{|\lambda_{a}|}{\Lambda} = 0.
\end{equation}
At $\omega = 0$ we have the following solution:
\begin{equation}
\lambda_{a} = \Lambda\exp[- \pi/Ng] \equiv Q_0. \label{eq:mass}
\end{equation}

The existence of the saddle point does not mean, however, that the system
has a non-zero order
parameter $\left<Q\right>$. Such order parameter in the present case
is the density of
states at the Fermi energy:
\begin{equation}
\rho(0) = \lim_{r \rightarrow 0}r^{-1}~
\frac{\partial Z}{\partial\omega}|_{\omega = 0}
= \lim_{r \rightarrow 0}\frac{1}{gr}~Tr(Q + Q^+).
\end{equation}
The appearence of the
average $\left<Q\right>$  breaks  the continuous symmetry $U(r)$, and
it is well known that such symmetry breaking cannot occur in two dimensional
systems due to strong transverse
fluctuations of $Q$. A finite $Re\left<Q\right>$ arises
only at finite $\omega$ when
the transverse
fluctuations acquire a finite correlation length
$\xi_c \sim \omega^{- 1/(2 - \Delta)}$.
$\Delta$ is the scaling dimension of the operator $Re Q$. In this case we have:
\begin{equation}
Re\left<Q\right> \sim Q_0~\omega^{\frac{\Delta}{2 - \Delta}}.
\label{eq:density}
\end{equation}

Now we are going to demonstrate that the described  qualitative picture
survives the replica
limit and maintains a contact with the previous treatment.
In order to show that the radial fluctuations of $Q$ remain
gapful even at $r = 0$, we use the exact solution of the model
(\ref{eq:nonabel}).
This exact solution was found by one of the authors \cite{Tsv87}. It
was shown that  the
model (\ref{eq:nonabel}) at $\omega = 0$
is a relativistic
limit of the model directly solvable by the Bethe ansatz:
\begin{eqnarray}
H = \int dx[ \psi_{a,\alpha}^+(- \frac{1}{2}\partial_x^2 -
\frac{1}{2}k_F^2)\psi_{a,\alpha} -
g~(\psi_{a,\alpha}^+\psi_{b,\alpha})(\psi_{b,\beta}^+\psi_{a,\beta})].
\label{eq:nonrel}
\end{eqnarray}
The basic features of the exact solution coincide with  those derived
from the non-Abelian bosonization.  In
particular, it was established in Ref.\onlinecite{Tsv87} that,
in the relativistic
limit, the spectrum of model
(\ref{eq:nonrel}) consists
of three sectors. Excitations of  one sector
are $U(r)$-singlets and have a spectral
gap given by Eq.(\ref{eq:mass}). The  low energy sector contains
gapless excitations which are $SU(N)$-singlets. One branch of gapless
excitations
consists of the U(1) scalar field described
by the effective action
\begin{equation}
S_1 = \frac{1}{2}\int d^2x (\partial_{\mu}\phi)^2 . \label{eq:bosons}
\end{equation}
while the other is described by the Wess-Zumino-Witten model
at the critical point,
i.e. by $H^0 _{SU(r)}$ in Eq.(\ref{eq:HamSug}). The latter
can be also written in the
Lagrangian form with the following action:
\begin{equation}
S_2 = \int d^2x \{\frac{N}{8\pi}Tr(\partial_{\mu}g^+\partial_{\mu}g) +
\frac{N}{24\pi^3}
\int_0^1 d\xi \epsilon_{abc}~
Tr(g^+\partial_a gg^+\partial_b gg^+\partial_c g)\}, \label{eq:wzw}
\end{equation}
where $g$ is a matrix from the $SU(r)$ group ($g^+g = I, \det g = 1$).
The second term in the r.h.s. of
Eq.(\ref{eq:wzw}) is the so-called  Wess-Zumino term. It has a topological
origin. Despite of the fact that
it is written as an integral including the
additional dimension, its actual value (modulus $2\pi iN$) depends on
the boundary values of $g(x,\xi = 0) = g(x)$ ($g(x,\xi = 1) = 0$). This
property follows from the fact that the Wess-Zumino term is proportional to
the
integral of the
Jacobian of transformation
from the three dimensional euclidean coordinates to the group coordinates,
thus being a total derivitive in disguise.
Excitations of the Wess-Zumino-Witten model
are gapless and the correlation functions are well studied
(see Ref. \onlinecite{Knizh84}). The presence of the Wess-Zumino term
is  absolutely crucial for the criticality - the model without it
undergoes a strong renormalization.

Let us establish a connection between the exact solution and  the semiclassical
approach. The massive sector obviously corresponds to radial fluctuations
of the order parameter field $Q$. The rest of the action (except of the
Wess-Zumino term)
represents the first term in the gradient expansion of the $Tr\ln$-term in
Eq.(\ref{eq:z}). Indeed,
at small momenta $|p| \ll Q_0$ this term is equal to
\begin{equation}
\frac{N}{8\pi Q_0^2}\int d^2x Tr(\partial_{\mu}Q^+\partial_{\mu}Q).
\label{eq:naive}
\end{equation}
Now let us write
down the $Q$-field in the form (\ref{eq:orderpar}).
 Substituting this expression
into Eq.(\ref{eq:naive}) and
taking into account that $Trg^+\partial_{\mu}g = 0$ as a trace
of an element of the
algebra,  we
reproduce Eq.(\ref{eq:bosons}) and the first term of Eq.(\ref{eq:wzw}).
It comes
not entirely  unexpected that the naive gradient expansion misses the important
Wess-Zumino term: the latter is
a Berry phase and requires special care. In order to avoid possible
calculational difficulties, we have resorted to the exact solution. In this
way we obtain the result having an independent value
(see also Ref.\onlinecite{Tsv94}):
\begin{eqnarray}
& &- Tr \; \ln[i\gamma_{\mu}\partial_{\mu} +
(1 + i\gamma_5)gQ_0/2 + (1 - i\gamma_5)g^+Q_0/2] = \nonumber\\
& & \int d^2x \{\frac{1}{8\pi}Tr(\partial_{\mu}g^+\partial_{\mu}g) +
\frac{1}{24\pi^3}\int_0^1 d\xi~ \epsilon_{abc}~
Tr(g^+\partial_a gg^+\partial_b gg^+\partial_c g)\} + O(Q_0^{-2}),
\end{eqnarray}
where $g$ belongs to $U(r)$ group ($r$ is now an arbitrary number not related
to the number of replicas).

The fact that radial and transverse fluctuations are decoupled
at the level of the
Bethe ansatz equations allows one to consider the corresponding replica limits
separately. The proof of the fact that the replica limit is  well defined for
the
gapful sector is given  in the Appendix. Here we  consider
only the replica limit for the
correlation functions of $Q$. For the pair correlation function we have
\begin{eqnarray}
<Q(x)Q^+(y)> &=& Q_0^2<\exp[ i \sqrt{4\pi/Nr}\phi(x)]
\exp[- i \sqrt{4\pi/Nr}\phi(y)]><g(x)g^+(y)> \nonumber\\
&\sim&
|x - y|^{ - 2 (1/Nr + \Delta_g)} = |x - y|^{ -(2/N^2)},
\end{eqnarray}
in agreement with the above estimation of the scaling dimension
of the operator $Q$
(see Eq.(\ref{eq:dim})).
According to Eq.(\ref{eq:density}) the dimension $1/N^2$  gives
the following estimate for the
density of states at $|\omega| \ll Q_0$:
\begin{equation}
\rho(\omega) \sim \omega^{1/(2N^2 - 1)}.
\end{equation}
In fact, even in the case $N = 2$ the corresponding
exponent is extremely small: $1/7$
and the
specific heat at low temperatures goes almost linearly: $c_v \sim T^{8/7}$.
\par
So far the impurity scattering between the nearest-neighbor nodes has been
neglected. When these processes are taken into account,
the effective replicated action can be easily shown to have the form:
\begin{equation}
S_{eff} = S_1 + S_2 + S_{12},
\label{eq:nnn}
\end{equation}
\begin{eqnarray}
S_1 &=& \int d^2 x \{ \bar{\eta}_{1a \alpha}
(v_1 \partial_1 \tau_1 + v_2 \partial_2 \tau_2
+ \omega_n)\eta_{1a \alpha}\nonumber\\
& &+ \frac{1}{2} g_3 (\bar{\eta}_{1a \alpha} \tau_1 \sigma^3 _{\alpha \beta}
\eta_{1a \beta})^2
+ \frac{1}{2} g_1 [(\bar{\eta \alpha}_{1a} \tau_2 \sigma^1 _{\alpha \beta}
\eta_{1a \beta})^2 +
(\bar{\eta \alpha}_{1a} \tau_2 \sigma^2 _{\alpha \beta} \eta_{1a \beta})^2
]\},\ \\
S_2 &=& \int d^2 x \{ \bar{\eta}_{1a \alpha}
(v_1 \partial_2 \tau_1 + v_2 \partial_1 \tau_2
+ \omega_n)\eta_{1a \alpha}\nonumber\\
& &+ \frac{1}{2} g_3 (\bar{\eta}_{2a \alpha} \tau_1
\sigma^3 _{\alpha \beta}\eta_{2a \beta})^2
+ \frac{1}{2} g_1 [(\bar{\eta}_{2a \alpha} \tau_2
\sigma^1_{\alpha \beta}\eta_{2a \beta})^2 +
(\bar{\eta}_{2a \alpha} \tau_2 \sigma^2 _{\alpha \beta}\eta_{2a \beta})^2 ]\},\
\\
S_{12} &=& \frac{1}{2} g_2 \sum_{i \neq j} \int d^2 x
\{ (\bar{\eta}_{ia \alpha} \tau_1 \sigma^3 _{\alpha \beta} \eta_{ja \beta})^2 -
(\bar{\eta}_{ia \alpha} \tau_1 \eta_{ja \alpha})^2\nonumber\\
& &+(\bar{\eta}_{ia \alpha} \tau_2 \sigma^2 _{\alpha \beta} \eta_{ja \beta})^2
+
(\bar{\eta}_{ia \alpha} \tau_2 \sigma^1 _{\alpha \beta}\eta_{ja \beta})^2 \},
\end{eqnarray}
where fermionic fields $\eta_1$ and $\eta_2$ describe two pairs of opposite
nodes,
$(1, \bar{1})$ and $(2, \bar{2})$, respectively.
The action (\ref{eq:nnn}) is still symmetric
under continuous $\gamma^5-$transformations
\begin{equation}
\eta_{ja} \rightarrow e^{i \alpha \tau_3}~\eta_{ja}, ~~~
\bar{\eta}_{ja} \rightarrow \bar{\eta}_{ja}~e^{i \alpha \tau_3}
\end{equation}
Therefore the DOS vanishes as $\omega \rightarrow 0$, even when
the nearest-neighbor internode scattering is included into consideration.
This conclusion does not depend on the ratio $v_1/v_2$. However, the
crossover from the two-node critical regime to the four-node one
will be shifted towards lower energies on increasing $v_1/v_2$.

\section{Nonmagnetic impurities in 2D orbital antiferromagnet}

In this section, we consider effects of nonmagnetic disorder in another 2D
electron system
with conical energy spectrum - the orbital antiferromagnet state (OAF).
This state was
earlier suggested
as one of possible phases of weakly interacting fermions on a square lattice
with a
half-filled energy band \cite{Ners88}. The OAF phase can be viewed as
resulting from
anomalous particle-hole pairing
$<c_{{\bf k},\alpha}^{+} c_{{\bf k}+{\bf Q},\alpha}>$
with momentum transfer equal to the nesting
vector ${\bf Q} = (\pi, \pi)$. Such a pairing breaks translational, point and
time
reversal
symmetries,
but preserves spin-rotational invariance. At the microscopical level, the OAF
is
characterized by nonzero local charge currents circulating around plaquettes in
an
alternating way. It actually represents a weak-coupling analog of the flux
phase
studied by Affleck and Marston \cite{Affleck88} in the large-$U$ Hubbard model.
\par
Since the OAF is a state with a spin-singlet particle-hole pairing, and since
we wish to discuss effects of nonmagnetic impurities, spin degrees of freedom
are
irrelevant and will be suppressed in what follows. Then
the mean-field Hamiltonian describing a 2D OAF state can be written as
\begin{equation}
H_{0} = \sum_{{\bf k}} [\epsilon({\bf k}) c^{+}_{{\bf k}}c_{{\bf k}}
- i\Delta({\bf k})c^{+}_{{\bf k}}c_{{\bf k}+{\bf Q}}],
\end{equation}
where the tight-binding spectrum $\epsilon({\bf k})$ and the gap function
$\Delta({\bf k})$
have the same two-cosine structures as those for the d-wave superconductor
(see (\ref{eq:Ham}))
with $\mu=0$.
The quasiparticle spectrum $E{({\bf k})} = \pm \sqrt
{\epsilon^{2}({\bf k}) + \Delta^{2}({\bf k})}$ has zeros at points
${\bf k}_{1} = -{\bf k}_{\bar{1}} = (\pi/2,\pi/2)$,
${\bf k}_{2} = -{\bf k}_{\bar{2}} = (-\pi/2,\pi/2)$. However, due to unit cell
doubling
taking place in the OAF state, the reduced Brillouin zone corresponding to the
new
translational
symmetry, coincides with the area enclosed by the original square Fermi surface
$ABCD$, shown in Fig. \ref{figure3}. Therefore,
there are actually $two$ unequivalent
nodes $1 = \bar{1}$ and $2 = \bar{2}$ in the OAF
phase, with two associated branches of low-energy excitations with conical
spectra.
\par
Assuming that ${\bf q}$ lies inside $ABCD$, we introduce a 2-spinor
\[
\Phi_{\bf q} = \left(
\begin{array}{c}
c_{\bf q}\\
c_{{\bf q}+{\bf Q}}
\end{array}
\right)
\]
and rewrite Hamiltonian (75) in a matrix form
\begin{equation}
\hat{H_{0}} = \sum_{{\bf q}}~ \Phi^{+}_{\bf q}~ [\epsilon({\bf q})
\hat{\tau_{3}}
+ \Delta({\bf q}) \hat{\tau_{2}}]~\Phi_{\bf q}.
\end{equation}
\par
Scattering on nonmagnetic impurities is decribed by
\begin{eqnarray}
H_{imp} &=& \sum_{{\bf k},{\bf k'}} V({\bf k}-{\bf k'}) c^{+}_{{\bf k}} c_{{\bf
k'}}
\nonumber\\
&=& \sum_{{\bf q},{\bf q'}}~ [V_{0}({\bf q}-{\bf q'})\Phi^{+}_{\bf q} \Phi_{\bf
q'}
+ V_{1}({\bf q}-{\bf q'}) \Phi^{+}_{\bf q} \hat{\tau_{1}} \Phi_{\bf q'}]
\end{eqnarray}
where $V({\bf k}-{\bf k'})$ is the Fourier transform of a Gaussian random
potential
defined on the original lattice. In the OAF state, this potential gives rise to
$two$ random fields, $V_{0}({\bf q}) = V({\bf q})$
 and
$V_{1}({\bf q}) = V({\bf q} + {\bf Q})$,
whose small-momentum parts describe
the chemical potential and charge-density-wave (CDW) fluctuations,
respectively.
\par
The physical difference between $V_{0}$ and $V_{1}$ allows to expect
qualitatively different effects
of these two types of disorder on the low-energy DOS in the OAF phase. Consider
a local,
slowly varying fluctuation of the chemical potential, such that $V_{0}(\bf x)$
is
nearly
constant in a macroscopically large region. The finite value of $V_{0}$ will
then
(locally) break the particle-hole symmetry of the OAF state and open a finite
Fermi surface,
thus resulting in a
finite DOS in the given region: $N_{loc} \sim |V_{0}|$.
The latter conclusion does not depend
on the magnitude and sign of $V_{0}(\bf x)$. Therefore, the DOS averaged over
realizations
of the field $V_{0}(\bf x)$ is expected to be finite at zero energy.
On the other hand, local CDW fluctuations tend to open
a gap in the excitation spectrum, so that,
for $V_{2}(\bf x) \simeq$ const, the local DOS will vanish in the whole energy
range
$|E| < |V_{2}|$. Therefore, we expect in this
case the averaged DOS to vanish at $E \rightarrow 0$
even faster than in the pure system. These expectations
will be confirmed by the analysis
presented below.
\par
Following the same strategy as in Sec.4, we pass to the continuum limit based
on
separating states near the nodes. For simplicity,
we shall ignore scattering between nodes
$1$ and $2$. We shall argue later that the internode
scattering does not alter main conclusions
obtained in this section. Specializing to the node $1$, we arrive at the
Euclidean action
\begin{equation}
S [V_i]  = \int d^{2}x \bar{\eta}(x) [ -i \partial_{1} \tau_{1}
- i \partial_{2} \tau_{2}
+ \mu (x) - m(x) \tau_{3} - i\omega_{n}] \eta(x) \label{eq:actionOAF}
\end{equation}
which
describes two-dimensional fermions
with a random mass $m(x) = V_{1}(x)$ and
chemical potential $\mu (x) = V_{0}(x)$.
The Gaussian
probability distributions for random fields $V_{i}(x),~(i = 1,2)$ are given by
\begin{equation}
P[V_{i}] = \int DV_{i} \exp~ [-\frac{1}{2g_{i}} \int V_{i}^{2}(x) d^{2}(x)],~~
(i = 0,1).
\end{equation}
\par
Comparing Eqs.(\ref{eq:actionOAF}) and (\ref{eq:action}), we observe the
important difference between
the d-wave superconductor and OAF.
In both cases we have a continuum description in terms of
2D "relativistic" fermions.
However, nonmagnetic impurities introduce different types of randomness to the
corresponding Dirac Hamiltonians: random gauge field in the former case, and
random mass and chemical potential in the latter case. This is related
to different symmetries of the condensates in the two cases.
\par
Averaging over disorder results in the following quantum
Hamiltonian of 1D interacting fermions:
\begin{equation}
H = \int dx [\eta^{+}_{a} (- i \partial_{x} \tau_{3} + \omega_{n} \tau_{2})
\eta_{a}
+ \frac{1}{2} g_{0} (:\eta^{+}_{a} \tau_{2} \eta_{a}:)^{2}
- \frac{1}{2} g_{1} (:\eta^{+}_{a} \tau_{1} \eta_{a}:)^{2}].
\label{eq:HamOAF}
\end{equation}
As before, $a = 1,2,...,r$ is the replica index.

At $\omega_n = 0$,
the case $g_{0} = - g_{1}$ would correspond to the massless chiral
Gross-Neveu model, whose symmetry group is $U(r) = U(1)\times SU(r)$.
However, since both coupling constants
are positive
by definition, this case is never realized; so even at $\omega_n = 0$
the model (\ref{eq:HamOAF}) has the $O(2r) = Z_2\times SO(2r)$ symmetry.
The important difference between these two groups is that the latter
does not contain the continuous subgroup $U(1)$ which survives in the
replica limit. Therefore it is possible that at $r = 0$ the model
 (\ref{eq:HamOAF}) has  no gapless excitations.
Thus the  symmetry can be spontaneously
broken, and contrary to the case of the
d-wave superconductor,
the DOS at $\omega_n = 0$, being the corresponding order parameter,
$\left<\bar{\eta}_a \eta_a\right>$, can be nonzero \cite{Fisher85}.
\par
The breakdown of the discrete symmetry must be signalled by
the development
of strong-coupling regime in the zero-replica limit.
To identify the effective couplings that are subject to renormalization,
we rewrite
the interaction terms in (\ref{eq:HamOAF}) as follows:
\begin{eqnarray}
H_{int} = - \frac{1}{4} \int dx ( \psi^{+}_{R,a} \psi^{+}_{R,b}
\tilde{\Gamma}_{aba'b'} \psi_{L,b'} \psi_{L,a'} + H.c. ) \nonumber\\
- \frac{1}{4} \int dx ( \psi^{+}_{R,a} \psi^{+}_{L,b}
\Gamma_{aba'b'} \psi_{L,b'} \psi_{R,a'} + H.c. ).
\end{eqnarray}
Here
\begin{eqnarray}
\tilde{\Gamma}_{aba'b'} &=&
\tilde{\Gamma} (\delta_{aa'} \delta_{bb'} - \delta_{ab'} \delta_{ba'}),
\nonumber\\
\Gamma_{aba'b'} &=&
\Gamma_2 \delta_{aa'} \delta_{bb'} - \Gamma_1 \delta_{ab'} \delta_{ba'}
\label{eq:coup}
\end{eqnarray}
are amplitudes of the Umklapp and momentum (chirality) conserving
scattering processes,
respectively. The bare values of the parameters in Eqs.(\ref{eq:coup}) are
\begin{equation}
\tilde{\Gamma}(0) = - (g_{0} + g_{1}), ~~
\Gamma_2(0) = - (g_{0} - g_{1}),~~\Gamma_1(0) = 0.
\end{equation}
In the leading logarithmic approximation (one-loop accuracy),
the renormalization group equations are
\begin{eqnarray}
\frac{\partial \tilde{\Gamma}}{\partial l}
&=& - [(r - 1) \Gamma_1 - 2 \Gamma_{2}] \tilde{\Gamma}, \nonumber\\
\frac{\partial \Gamma_1}{\partial l}
&=& - \frac{1}{2}[r \Gamma_{1}^{2} + (r - 2) \tilde{\Gamma}^{2}],\nonumber\\
\frac{\partial \Gamma_2}{\partial l}
&=& \frac{1}{2} (\tilde{\Gamma}^{2} - \Gamma_1^{2}),
\label{eq:RG}
\end{eqnarray}
where $l = \ln (\Lambda/|\epsilon|)$ is a logarithmic variable.

It is instructive to consider first a somewhat artificial case, when only
backward
impurity scattering,
corresponding to a random Dirac mass $m(x)$ in Eq.(\ref{eq:actionOAF}), is
present,
$g_{0} = 0,~ g_{1} \neq 0$.
The solution of Eqs.(\ref{eq:RG})
\begin{equation}
\Gamma_{2} = 0, ~~~\Gamma_1(l) = \tilde{\Gamma}(l) =
- \frac{g_{1}}{1 - (r - 1)g_{1} l}
\end{equation}
indicates that a strong-coupling regime, taking place at all $r > 1$, is
changed
by a weak-coupling ("zero-charge") infrared behavior in the replica limit:
\begin{equation}
\Gamma_1(l; r \rightarrow 0)) \simeq  - \frac{1}{\ln~(\Lambda/|\epsilon|)}.
\end{equation}
Therefore, in this case the perturbation
caused by impurities is marginally irrelevant, the discrete $\gamma^5$-symmetry
remains unbroken, and the DOS will show a power-law energy dependence.

On the other hand, when only forward scattering is considered (random chemical
potential $\mu(x)$),  $g_{1} = 0,~ g_{0} \neq 0$,
the solution of Eqs.(84) is
\begin{equation}
\Gamma_{2} = 0, ~~~\Gamma_1(l) = - \tilde{\Gamma}(l) =
\frac{g_{0}}{1 + (r - 1)g_{0} l}
\end{equation}
and the situation is inverted. The weak-coupling regime, taking place at $r >
1$,
is changed by a strong-coupling one at $r \rightarrow 0$,
signalling breakdown of the discrete $\gamma^5$-symmetry.
The position of the pole
of the amplitude $\Gamma_1(l; r \rightarrow 0)$ determines the
 correlation length of the system in the zero-energy limit,
$\xi_c^{-1} \sim \Lambda \exp (-1/g_{0})$.
As a result, the DOS
at $\omega = 0$ is finite.

It can be shown that, when both scattering processes are present, the solution
of
renormalization group equations (\ref{eq:RG}) is always singular. This
clarifies
the special role of random chemical potential which drives the system
towards strong coupling regime with a finite DOS
at arbitrarily small $g_{0}$. Therefore,
as long as $g_{0} \neq 0$, taking into account
the scattering between the nodes 1 and 2
would not qualitatively change this conclusion.

\section{Conclusions}

 In this paper we have presented a detailed analysis of several
two-dimensional models with quenched randomness. The obtained results
are essentially non-perturbative and, apart from their practical
significance, give us a new  understanding of the theoretical side
of the problem.

We have shown that, for weak nonmagnetic impurities in a 2D d-wave
superconductor,
the whole diagrammatic series for the single-electron self energy has to be
summed up.  By explicit calculations and symmetry arguments, we have
demonstrated that the DOS $\rho(\omega)$, averaged over disorder, preserves the
property of being zero at zero energy. A new power-law behavior
$\rho(\omega) \sim |\omega|^{\alpha}$ was found, with critical exponent
$\alpha$
depending on the type of included scattering processes.

The feature most  essential for our solution  is
that our  models  have a relativistic
spectrum. This fact allows us
to use a powerful machinery of Abelian and
non-Abelian bosonization and the Bethe ansatz. Some of the models
turned out to be exactly solvable and the main technical problem was
to derive the replica limit of this exact solution.
On the basis of our experience we can
formulate a principle that the right procedure is to deal not with
the energy eigenvalues, but with correlation functions. Energy
eigenvalues in different symmetry sectors are singular at $r
\rightarrow 0$, but  these singularities cancel in correlation
functions.

\acknowledgments

We acknowledge the generous hospitality of the
International Centre for Theoretical Physics in Trieste (Italy),
where this collaboration was started.
A.N. and F.W. acknowledge the financial
support from Chalmers University of Technology and
would like to thank H. Johannesson, S. \"{O}stlund, B.-S. Skagerstam
and T. Einarsson for
helpful discussions. A.N. is grateful to A. Luther for his generous
hospitality at Nordita and interesting conversations. A.T. is
grateful to J. Chalker and B. Altshuler for their interest to
the work.

\appendix
\section*{}
 Let us consider the thermodynamic Bethe ansatz
equations derived for the
massive sector of the model (\ref{eq:nonrel}) in \onlinecite{Tsv87}.
For the sake of simplicity we consider only the case $N = 2$. The free
energy is described by the following system of nonlinear integral equations:
\begin{eqnarray}
\frac{F_r}{L} &=& - \frac{Q_0}{4}\int dx \cosh (\pi x/2)\ln\{1 +
\exp[\beta\epsilon_r(x)]\}, \label{eq:free}\\
\beta\epsilon_n &=& s*\ln\{1 + \exp[\beta\epsilon_{n - 1}(x)]\}\{1 +
\exp[\beta\epsilon_{n + 1}(x)]\} \nonumber\\
& &- \beta Q_0 \cosh (\pi x/2)\delta_{n,r}, \label{eq:bethe}
\end{eqnarray}
where $n = 1,2, ...$ and
\[
s*f(x) = \int_{-\infty}^{\infty}dy \frac{f(y)}{4\cosh [\pi(x - y)/2]}.
\]
In the replica limit we need to show that the following quantity has a finite
value:
\begin{equation}
F = \lim_{r \rightarrow 0}\frac{F_r}{r}.
\end{equation}
It what follows we shall consider two limiting cases: $\beta Q_0 \ll 1$
and $\beta Q_0 \gg 1$.
In the first case, the main contribution to the free energy comes from large
$|x|$ where one can neglect the spectral gap and make the following
approximation:
\begin{equation}
 Q_0 \cosh (\pi x/2) \approx \exp( - \pi |x|/2 + \ln Q_0).
\end{equation}
The free energy at such $\beta$ is a free energy of a theory with linear
spectrum. It has been shown that this theory is the Wess-Zumino-Witten
model on the $SU(2)$ group. The free energy was calculated in Refs.
\onlinecite{Tsv85} and \onlinecite{Affleck86}. :
\begin{equation}
\frac{F_r}{L} = - \frac{\pi}{6} \frac{3r}{r + 2}~\beta^{-2}.
\end{equation}
The replica limit exists:
\begin{equation}
\frac{F}{L} = - \frac{\pi}{4}~\beta^{-2}.
\end{equation}
In the second case, we shall expand $F_r$ order by order in
$\exp( - \beta Q_0)$ and calculate the replica limit of this expansion.
The existence of the replica limit at $\beta Q_0 \ll 1$ guarantees a
convergence of this expansion.
In the first approximation at $\beta Q_0 \gg 1$
one can substitute $\epsilon_r = - \infty$ in
Eqs.(\ref{eq:bethe}). The solution for $n \neq r$ has the following form
\cite{Tsv85}:
\begin{eqnarray}
 1 + \exp[\beta\epsilon_{n}(x)] = \left[ \begin{array}{cc}
(n - r + 1)^2 & ( n > r)\\
\left(\frac{\sin\frac{\pi(n + 1)}{r + 2}}{\sin\frac{\pi}{r + 2}}\right)^2 & (n
< r).
\end{array}
\right .
\end{eqnarray}
Substituting $\epsilon_{r \pm 1}$ into the equation for $\epsilon_r$ we get
the second iteration:
\begin{equation}
\beta\epsilon_r^{(2)} = - \beta Q_0 \cosh (\pi x/2) +
\ln~[2~\frac{\sin(\pi r/r + 2)}{\sin(\pi/r + 2)}].
\end{equation}
Substituting it into Eq.(~\ref{eq:free}) and taking the replica limit, we
get the first iteration for the free energy:
\begin{eqnarray}
\frac{F^{(1)}}{L} &=& - \beta^{-1}\frac{\pi Q_0}{4}\int dx \cosh (\pi x/2)
\exp[ - Q_0\beta \cosh (\pi x/2)] \nonumber\\
&=& - \frac{1}{2}\int_{- \infty}^{\infty} dp
\exp[- \beta\sqrt{p^2 + Q_0^2}].
\end{eqnarray}
The first iteration is well defined and contains the scale $Q_0$. It is
possible
to calculate the next term in our expansion. To do this we solve
Eqs.(~\ref{eq:bethe}) linearizing them in
\[
d_n(x) = \ln\{1 + \exp[\beta\epsilon_n(x)]\}
- \ln\{1 + \exp[\beta\epsilon_n^{(0)}]\}.
\]
A similar procedure was applied in Ref. \onlinecite{Tsv872} for
the O(3) nonlinear sigma model. Taking the corresponding solutions from
Appendix of Ref. \onlinecite{Tsv872} and substituting $r = 0$ we get
the third iteration for $\epsilon_r$ at $r \rightarrow 0$:
\begin{eqnarray}
\beta\epsilon_r^{(3)}(x) &=&  - \beta Q_0 \cosh (\pi x/2) + \ln(\pi r)
\nonumber\\
& & -\frac{1}{2}\int dy \frac{1}{\cosh^2[\pi(x - y)/4]}\exp[ - \beta Q_0
\cosh (\pi y/2)].
\end{eqnarray}
 From here we get the following expression for the second
iteration of the free
energy:
\begin{equation}
\frac{F^{(2)}}{L} = \beta^{-1}\frac{\pi Q_0}{8}\int dxdy
\exp[ - Q_0\cosh (\pi x/2) - Q_0\cosh (\pi y/2)]
\frac{\cosh \pi(x + y)/4}{\cosh \pi(x - y)/4}.
\end{equation}

 Thus we see that the procedure does work and the $r = 0$ limit
exists.

\begin{figure}
\caption{Brillouin zone with the four nodes in the quasiparticle
spectrum of the 2D $d$-wave
superconductor labeled by $1, \bar 1, 2, \bar 2$. Typical
lines of constant quasiparticle energy are indicated.}
\label{figure1}
\end{figure}
\begin{figure}
\caption{The standard diagrammatic expansion
of the self-energy $\Sigma(k,\omega_n)$
for weak static non-magnetic impurity
scattering. The full line stands for the propagation of a quasiparticle
with a certain momentum and each broken semi-circle corresponds to the
Born scattering at the same impurity center. The frequency $\omega_n$ is
conserved
and therefore omitted.}
\label{figure2}
\end{figure}
\begin{figure}
\caption{The two nodal points for the  2D orbital antiferromagnet
(OAF). The Fermi surface of the underlying half-filled
tight-binding band is the
square $ABCD$ which is also the true Brillouin zone for the OAF.
Therefore there are only two independent nodes $1,2$ in the OAF.}
\label{figure3}
\end{figure}

\end{document}